\documentclass[letterpaper,12pt]{article} 
\usepackage{osameet3} 
\usepackage{ifthen} 
\newboolean{pdflatex}
\setboolean{pdflatex}{true} 

\newboolean{articletitles}
\setboolean{articletitles}{true} 

\newboolean{uprightparticles}
\setboolean{uprightparticles}{false} 

\newboolean{inbibliography}
\setboolean{inbibliography}{false} 
\def\paperauthors{LHCb collaboration} 
\def\paperasciititle{JPsi UPC} 
\def\paperkeywords{{High Energy Physics}, {LHCb}} 
\def\papercopyright{CERN on behalf of the LHCb collaboration}
\def\papercopyright{\the\year\ CERN for the benefit of the LHCb collaboration} 

\def\paperlicenceurl{https://creativecommons.org/licenses/by/4.0/}

\usepackage[top=1in, bottom=1.25in, left=1in, right=1in]{}
\usepackage{microtype}
\usepackage{lineno}  
\usepackage{xspace} 
\usepackage{caption} 

\usepackage{graphicx}  
\usepackage{color}
\usepackage{colortbl}
\graphicspath{{./figs/}} 
\DeclareGraphicsExtensions{.pdf,.PDF,png,.PNG}

\usepackage{amsmath} 
\usepackage{amssymb}
\usepackage{amsfonts}
\usepackage{upgreek} 

\newcommand*\patchAmsMathEnvironmentForLineno[1]{%
\expandafter\let\csname old#1\expandafter\endcsname\csname #1\endcsname
\expandafter\let\csname oldend#1\expandafter\endcsname\csname
end#1\endcsname
 \renewenvironment{#1}%
   {\linenomath\csname old#1\endcsname}%
   {\csname oldend#1\endcsname\endlinenomath}%
}
\newcommand*\patchBothAmsMathEnvironmentsForLineno[1]{%
  \patchAmsMathEnvironmentForLineno{#1}%
  \patchAmsMathEnvironmentForLineno{#1*}%
}
\AtBeginDocument{%
\patchBothAmsMathEnvironmentsForLineno{equation}%
\patchBothAmsMathEnvironmentsForLineno{align}%
\patchBothAmsMathEnvironmentsForLineno{flalign}%
\patchBothAmsMathEnvironmentsForLineno{alignat}%
\patchBothAmsMathEnvironmentsForLineno{gather}%
\patchBothAmsMathEnvironmentsForLineno{multline}%
\patchBothAmsMathEnvironmentsForLineno{eqnarray}%
}

\usepackage{hyperxmp}

\usepackage[pdftex,
            pdfauthor={\paperauthors},
            pdftitle={\paperasciititle},
            pdfkeywords={\paperkeywords},
            pdfcopyright={Copyright (C) \papercopyright},
            pdflicenseurl={\paperlicenceurl}]{hyperref}

\usepackage[colorinlistoftodos,textsize=scriptsize]{todonotes}

\usepackage[all]{hypcap} 

\usepackage{xspace} 
\usepackage{upgreek}

\def\logpt2 {\ensuremath{\ln(\pt^{*2})}\xspace}   
\def\lhcb   {\mbox{LHCb}\xspace}

\def\alice  {\mbox{ALICE}\xspace}

\def\lhc    {\mbox{LHC}\xspace}

\def\MagUp {\mbox{\em Mag\kern -0.05em Up}\xspace}

\ifthenelse{\boolean{uprightparticles}}%
{

 \def\Pmu         {\ensuremath{\upmu}\xspace}

 \def\Ppsi        {\ensuremath{\uppsi}\xspace}

 \def\PDelta      {\ensuremath{\Delta}\xspace}                 
 \def\PXi         {\ensuremath{\Xi}\xspace}                 
 \def\PLambda     {\ensuremath{\Lambda}\xspace}                 
 \def\PSigma      {\ensuremath{\Sigma}\xspace}                 
 \def\POmega      {\ensuremath{\Omega}\xspace}                 
 \def\PUpsilon    {\ensuremath{\Upsilon}\xspace}

 \def\PB      {\ensuremath{\mathrm{B}}\xspace}                 
                  
 \def\PD      {\ensuremath{\mathrm{D}}\xspace}

 \def\PJ      {\ensuremath{\mathrm{J}}\xspace}                 
 \def\PK      {\ensuremath{\mathrm{K}}\xspace}

 \def\Pi      {\ensuremath{\mathrm{i}}\xspace}

 \def\Ps      {\ensuremath{\mathrm{s}}\xspace}

 \def\thebaroffset{0.0em}
}
{

 \def\Pmu         {\ensuremath{\mu}\xspace}

 \def\Ppsi        {\ensuremath{\psi}\xspace}                 
                  
 \mathchardef\PDelta="7101
 \mathchardef\PXi="7104
 \mathchardef\PLambda="7103
 \mathchardef\PSigma="7106
 \mathchardef\POmega="710A
 \mathchardef\PUpsilon="7107
                  
 \def\PB      {\ensuremath{B}\xspace}                 
                  
 \def\PD      {\ensuremath{D}\xspace}

 \def\PJ      {\ensuremath{J}\xspace}                 
 \def\PK      {\ensuremath{K}\xspace}

 \def\Pi      {\ensuremath{i}\xspace}

 \def\Ps      {\ensuremath{s}\xspace}

 \def\thebaroffset{0.18em}
}
\newcommand{\offsetoverline}[2][\thebaroffset]{\kern #1\overline{\kern -#1 #2}}%

\makeatletter
\ifcase \@ptsize \relax
  \newcommand{\miniscule}{\@setfontsize\miniscule{4}{5}}
\or
  \newcommand{\miniscule}{\@setfontsize\miniscule{5}{6}}
\or
  \newcommand{\miniscule}{\@setfontsize\miniscule{5}{6}}
\fi
\makeatother

\DeclareRobustCommand{\optbar}[1]{\shortstack{{\miniscule (\rule[.5ex]{1.25em}{.18mm})}
  \\ [-.7ex] $#1$}}

\def\mumu       {{\ensuremath{\Pmu^+\Pmu^-}}\xspace}

\def\squark    {{\ensuremath{\Ps}}\xspace}

\def\KorKbar {\kern \thebaroffset\optbar{\kern -\thebaroffset \PK}{}\xspace}

\def\DorDbar {\kern \thebaroffset\optbar{\kern -\thebaroffset \PD}\xspace}

\def\B       {{\ensuremath{\PB}}\xspace}

\def\BorBbar {\kern \thebaroffset\optbar{\kern -\thebaroffset \PB}\xspace}

\def\Bd      {{\ensuremath{\B^0}}\xspace}

\def\BdorBdbar {\kern \thebaroffset\optbar{\kern -\thebaroffset \Bd}\xspace}

\def\Bs      {{\ensuremath{\B^0_\squark}}\xspace}

\def\BsorBsbar {\kern \thebaroffset\optbar{\kern -\thebaroffset \Bs}\xspace}

\def\jpsi     {{\ensuremath{{\PJ\mskip -3mu/\mskip -2mu\Ppsi\mskip 2mu}}}\xspace}
\def\psitwos  {{\ensuremath{\Ppsi{(2S)}}}\xspace}

\def\Y#1S{\ensuremath{\PUpsilon{(#1S)}}\xspace}

\def\LorLbar     {\kern \thebaroffset\optbar{\kern -\thebaroffset \PLambda}\xspace}

\def\AT#1     {\ensuremath{A_{\mathrm{T}}^{#1}}\xspace}           

\def\C#1      {\ensuremath{\mathcal{C}_{#1}}\xspace}                       
\def\Cp#1     {\ensuremath{\mathcal{C}_{#1}^{'}}\xspace}                    
\def\Ceff#1   {\ensuremath{\mathcal{C}_{#1}^{\mathrm{(eff)}}}\xspace}        
\def\Cpeff#1  {\ensuremath{\mathcal{C}_{#1}^{'\mathrm{(eff)}}}\xspace}       
\def\Ope#1    {\ensuremath{\mathcal{O}_{#1}}\xspace}                       
\def\Opep#1   {\ensuremath{\mathcal{O}_{#1}^{'}}\xspace}                    

\newcommand{\aunit}[1]{\ensuremath{\text{\,#1}}}       

\newcommand{\tev}{\aunit{Te\kern -0.1em V}\xspace}
\newcommand{\gev}{\aunit{Ge\kern -0.1em V}\xspace}
\newcommand{\mev}{\aunit{Me\kern -0.1em V}\xspace}
\newcommand{\kev}{\aunit{ke\kern -0.1em V}\xspace}
\newcommand{\ev}{\aunit{e\kern -0.1em V}\xspace}
\newcommand{\mevc}{\ensuremath{\aunit{Me\kern -0.1em V\!/}c}\xspace}
\newcommand{\gevc}{\ensuremath{\aunit{Ge\kern -0.1em V\!/}c}\xspace}
\newcommand{\mevcc}{\ensuremath{\aunit{Me\kern -0.1em V\!/}c^2}\xspace}
\newcommand{\gevcc}{\ensuremath{\aunit{Ge\kern -0.1em V\!/}c^2}\xspace}

\def\gsim{{~\raise.15em\hbox{$>$}\kern-.85em
          \lower.35em\hbox{$\sim$}~}\xspace}
\def\lsim{{~\raise.15em\hbox{$<$}\kern-.85em
          \lower.35em\hbox{$\sim$}~}\xspace}

\def\pt         {\ensuremath{p_{\mathrm{T}}}\xspace}

\def\tell1  {TELL1\xspace}
\def\ukl1   {UKL1\xspace}

\usepackage{cite} 
\usepackage{mciteplus}

\usepackage{longtable} 
\usepackage[bottom]{footmisc}
\begin{document}
\title{Quarkonia production in (ultra-)peripheral PbPb collisions at LHCb}


\author{Xiaolin Wang on behalf of the LHCb collaboration \\ xiaolin.wang@cern.ch}
\address{South China Normal University, Guangzhou, China}
\email{Presented at DIS2022: XXIX International Workshop on Deep-Inelastic Scattering and Related Subjects, Santiago de Compostela, Spain, May 2-6 2022.} 

\begin{abstract}
The cross-sections of coherent \jpsi and \psitwos production in ultra-peripheral PbPb collisions at a nucleon-nucleon centre-of-mass energy of $5.02\,\mathrm{TeV}$ are measured using a data sample collected in 2018 at $\lhcb$, and the differential cross-sections are measured separately as a function of transverse momentum and rapidity. The photo-production of \jpsi mesons at low transverse momentum is studied in peripheral PbPb collisions, \jpsi candidates are reconstructed through the prompt decay into \mumu in the rapidity region of $2.0<y<4.5$. These results significantly improve previous measurements and are compared to the latest theoretical predictions.
\end{abstract}
\section{Introduction}
The $\lhcb$ detector is a single-arm forward spectrometer covering the unique \mbox{pseudorapidity} range $2<\eta <5$~\cite{LHCb-DP-2008-001,LHCb-DP-2014-002}, 
which could provide precise vertex reconstruction, high particle momentum resolution and excellent particle identification ability. 
The quarkonia production in (ultra-)peripheral collisions is sensitive to the gluon parton distribution function in nuclear at small-x, and coherent photo-production would provide an excellent laboratory to study the nuclear shadowing effects and the initial state of quark-gluon plasma with small-x at the \lhc~\cite{Jones:2015nna}. 

\section{Study of charmonium production in ultra-peripheral PbPb collisions at LHCb~\cite{LHCb:2022ahs}}
Ultra-peripheral collisions(UPCs) occur when the distance between the center of two nuclei is larger than the sum of their radii~\cite{Bertulani:2005ru}. The two ions interact via their cloud of semi-real photons and photon-nuclear interactions dominate. The basic process is electromagnetic and strong interactions are suppressed. 
In UPCs, \jpsi and \psitwos mesons are produced from the colourless exchange of a photon from one nucleus and a pomeron from the other.
Coherent production occurs when the photon interacts with a pomeron emitted by the entire nucleus, which is the process we are going to study.

\begin{figure}[htbp]
\begin{center}
\includegraphics[width=0.50\linewidth, page={1}]{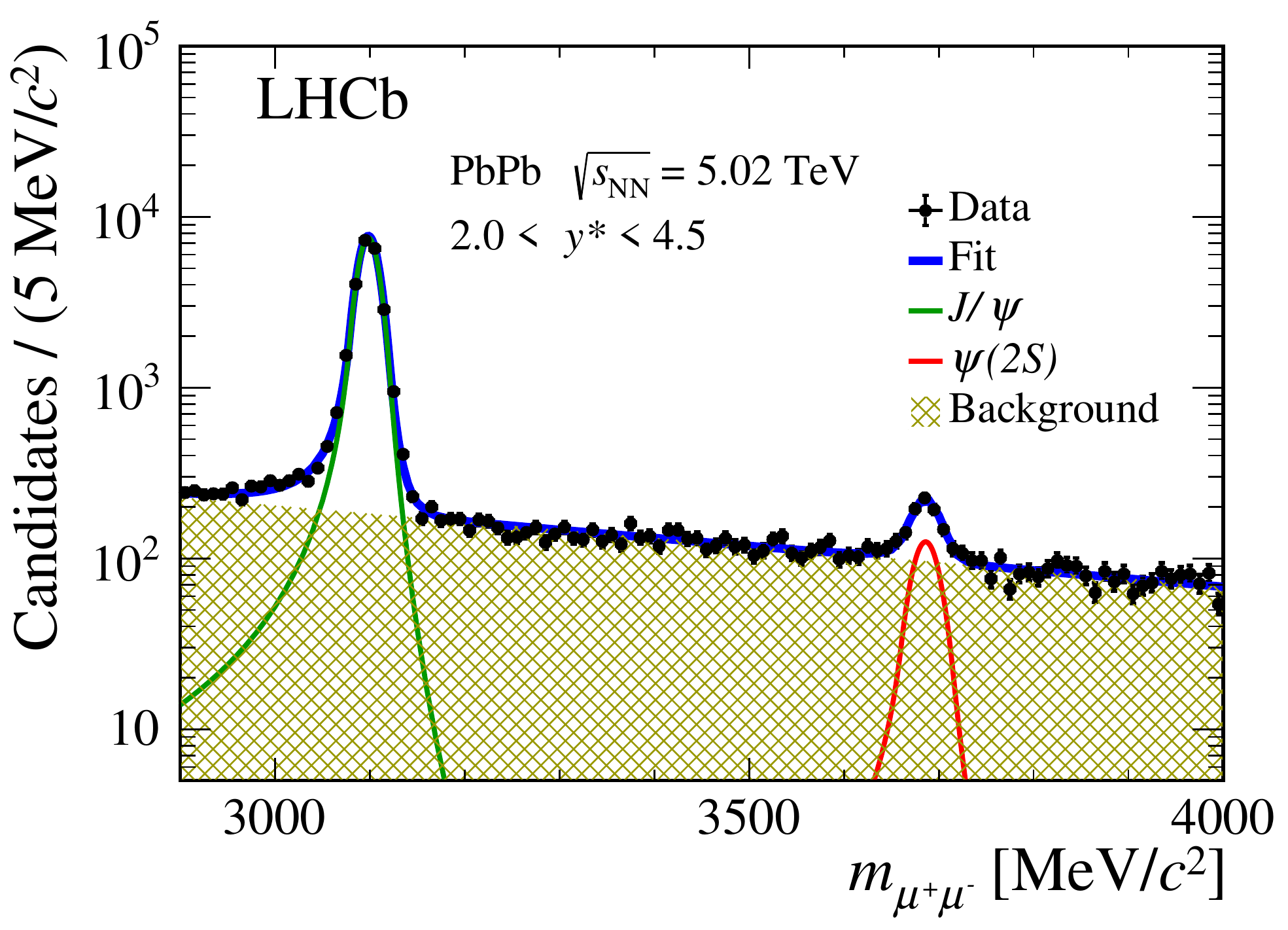}
\end{center}
\caption{Dimuon mass distribution fit for signal candidates in the rapidity range $2.0<y^{*}<4.5$.}
\label{fig:massfit}
\end{figure}

\begin{figure}[htbp]
    \centering
    \hfil
    \begin{minipage}[t]{0.49\linewidth}
        \centering
        \includegraphics[width=\linewidth, page = {1}]{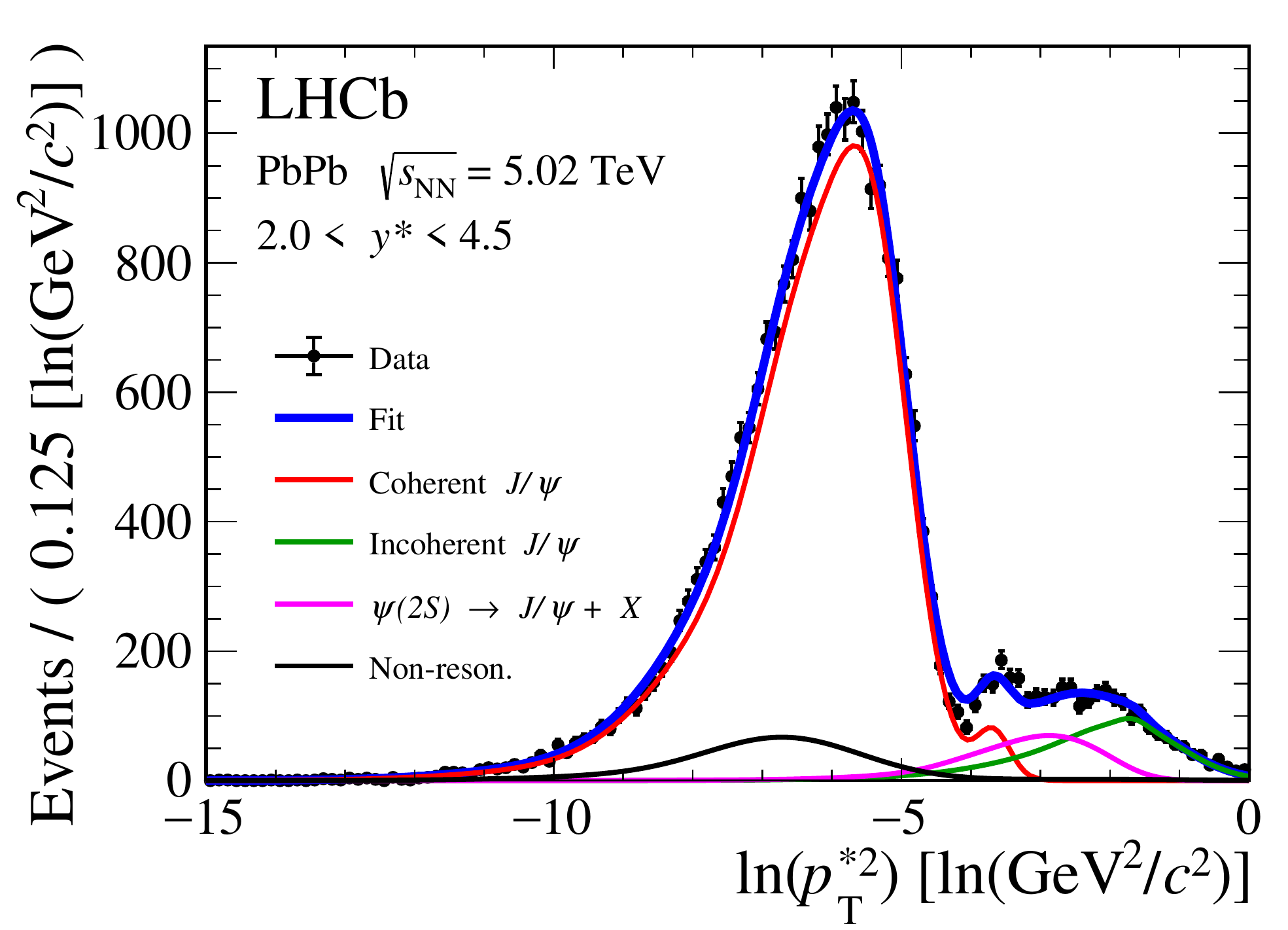}
        \put(-40,125){\jpsi}
    \end{minipage}
    \begin{minipage}[t]{0.49\linewidth}
        \centering
        \includegraphics[width=\linewidth, page = {1}]{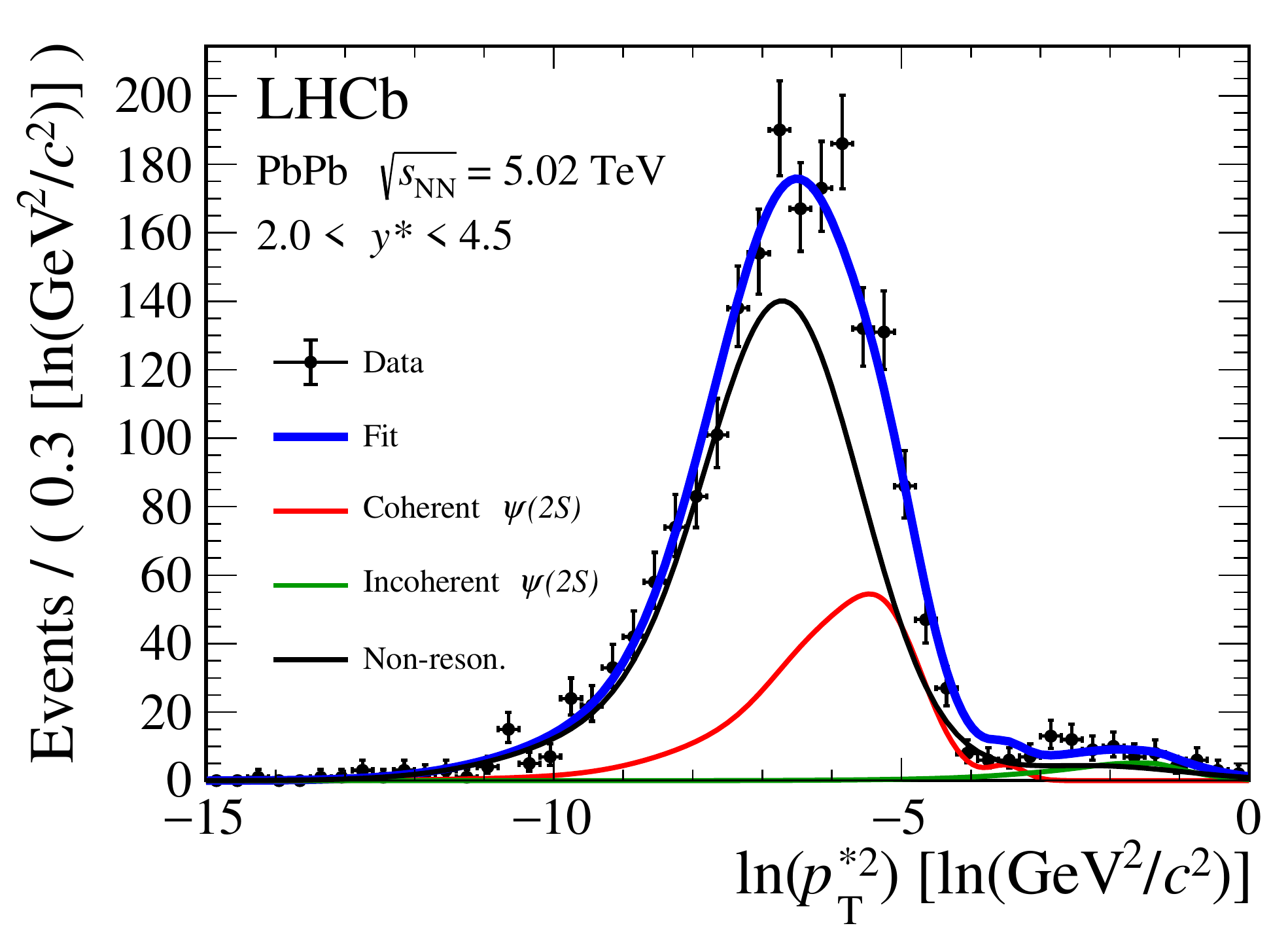}
        \put(-50,125){\psitwos}
    \end{minipage}
    \hfil
    \caption{The \logpt2 distribution fit of dimuon candidates in the $2.0<y^{*}<4.5$ range
    for (left) \jpsi candidates and (right) 
    \psitwos candidates.}
    \label{fig:2d}
\end{figure}

The charmonium candidates are reconstructed through the $\jpsi\rightarrow\mumu$ and $\psitwos\rightarrow\mumu$ decay channels, using 2018 PbPb data sample corresponding to an integrated luminosity of $228\pm10\,\mathrm{\mu b}^{-1}$. 
The signal yields are extracted in two steps.
First, a fit on the dimuon invariant mass spectrum is needed to estimate the charmonium and non-resonant background yields within the \jpsi and \psitwos mass windows, respectively. The mass distribution along with the dimuon mass fit is shown in Fig~\ref{fig:massfit}. 
Then fits to \jpsi or \psitwos \logpt2 distributions are performed to determine the coherent production events from the yields in corresponding mass windows, where the starred notation indicates that the observable is defined in the nucleus-nucleus centre-of-mass frame. 
Figure~\ref{fig:2d} shows the fits to the \logpt2 distributions of selected \jpsi and \psitwos candidates within rapidity interval $2.0<y^{*}<4.5$.
The shapes of coherent, incoherent and \psitwos feed-down components are taken from simulation, and the shape of non-resonant background is taken from 
the mass window, 
$3.2 < m_{\mumu} < 3.6 \gevcc$, 
where there is no signal component. 
The yields of the non-resonant background 
are determined as the integral of the non-resonant
component from the dimuon mass fit in
the \jpsi and \psitwos mass windows. 

\begin{figure}[htbp]
    \centering
    \hfil
    \begin{minipage}[t]{0.295\linewidth}
        \centering
        \includegraphics[width=\linewidth]{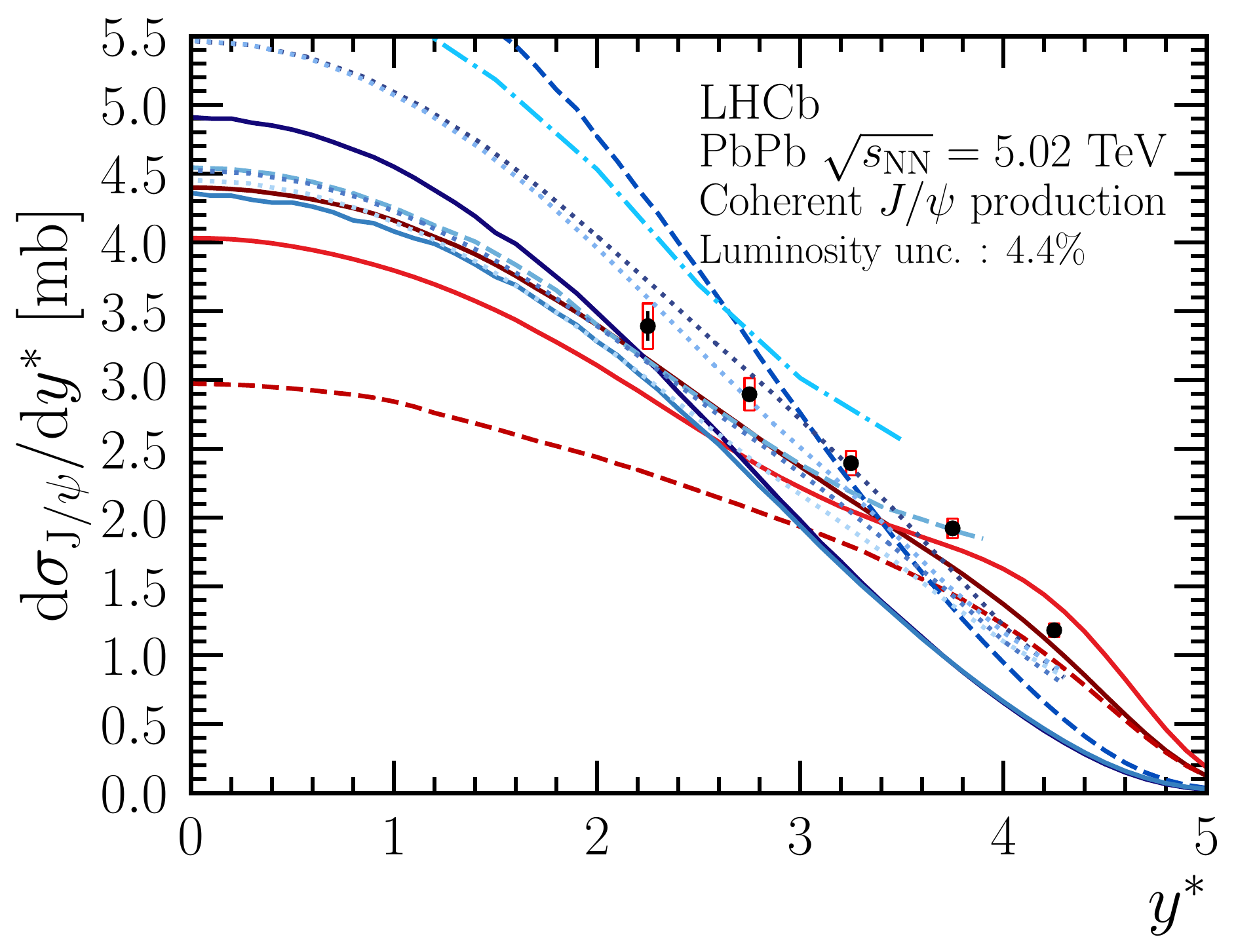}
    \end{minipage}
    \begin{minipage}[t]{0.295\linewidth}
        \centering
        \includegraphics[width=\linewidth]{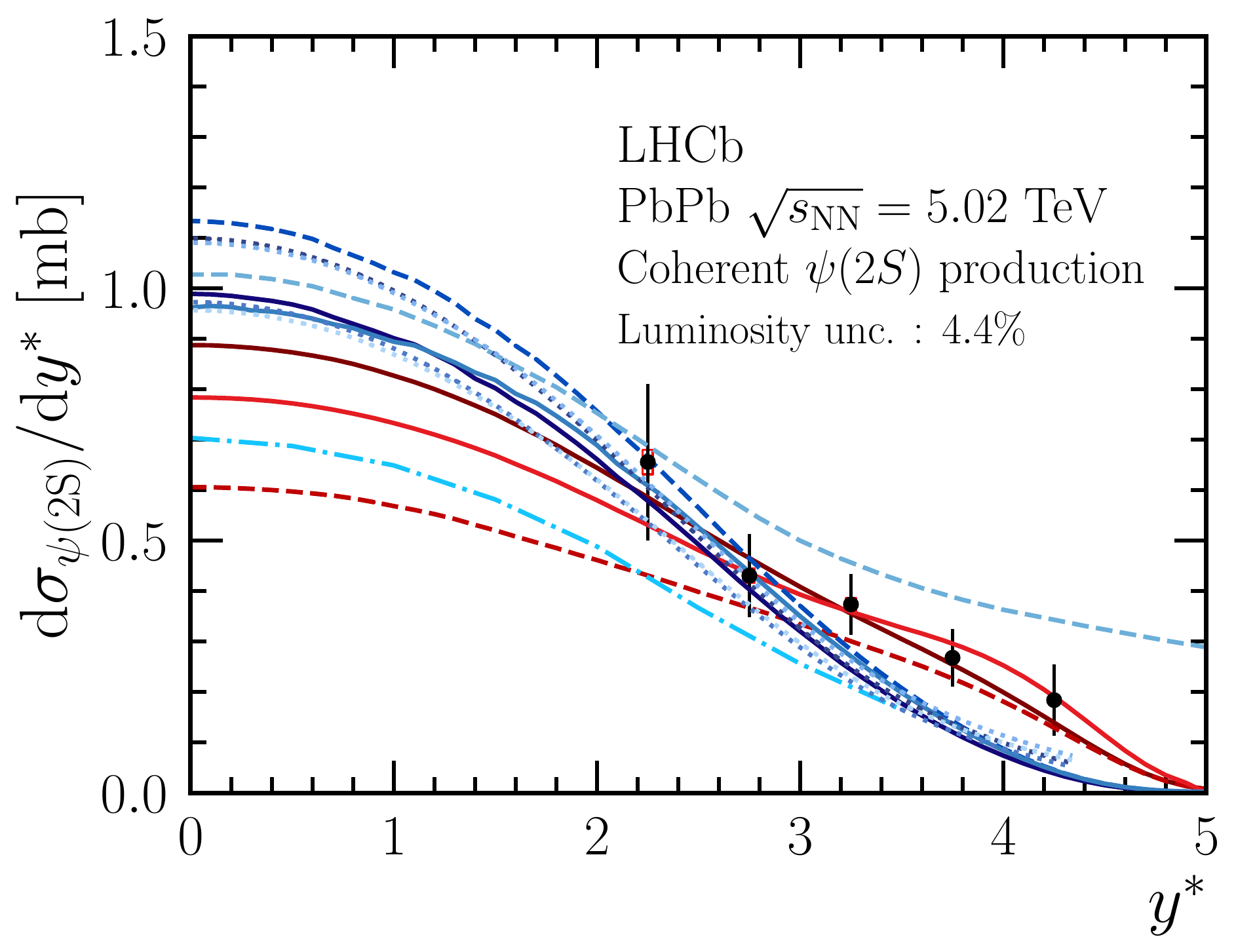}
     \end{minipage}
     \begin{minipage}[t]{0.39\linewidth}
        \centering
        \includegraphics[width=\linewidth]{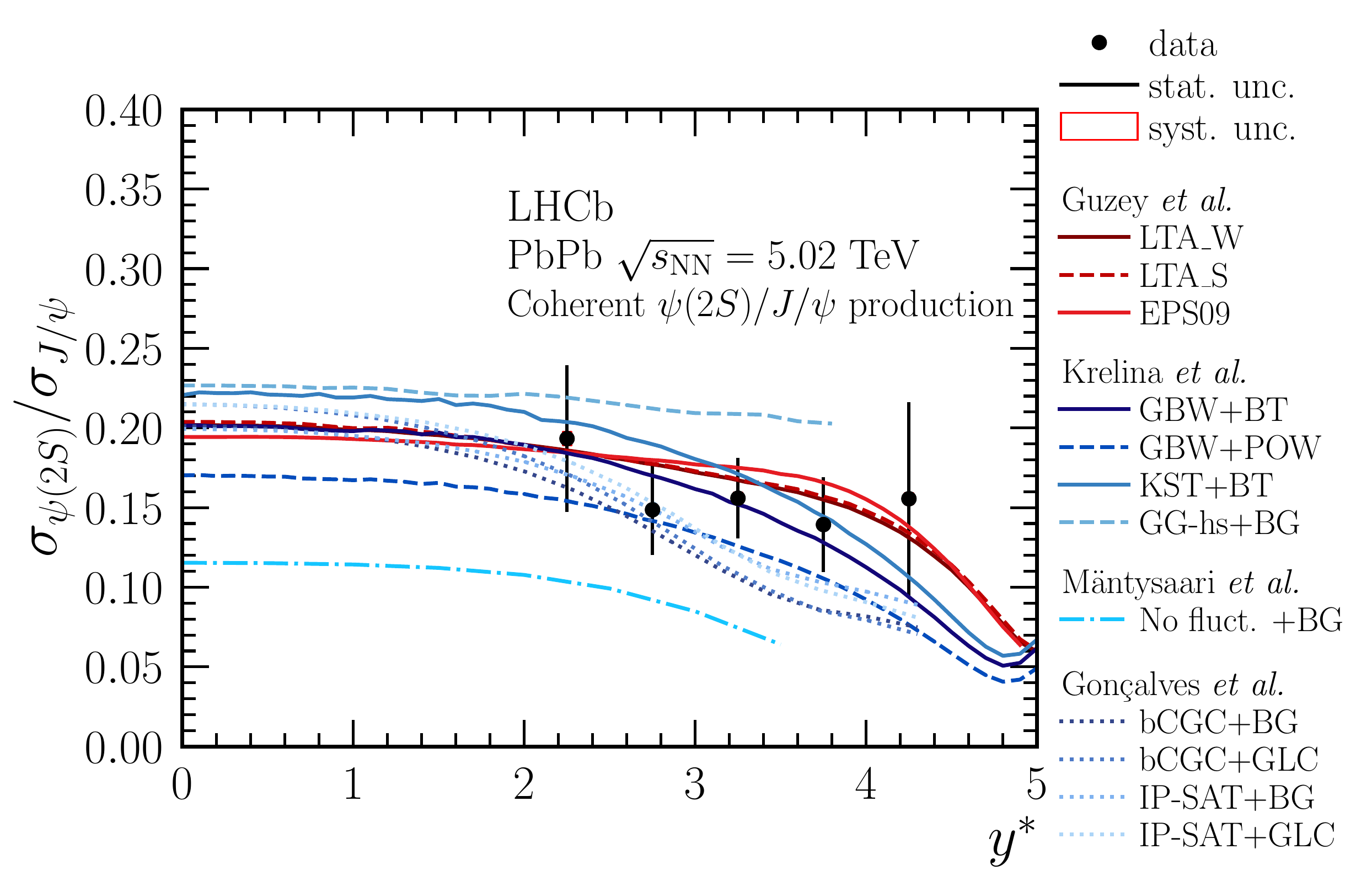}
     \end{minipage}
     \hfil
    \caption{Differential cross-section as a function $y^*$ for coherent $\jpsi$ , \psitwos production and the ratio of coherent \psitwos to \jpsi production, compared to theoretical predictions. These models are grouped as (red lines) perturbative-QCD calculations and (blue lines) colour-glass-condensate models.}
    \label{fig:theo_y}
\end{figure}

\begin{figure}[t]
    \centering
    \hfil
    \begin{minipage}[t]{0.4\linewidth}
        \centering
        \includegraphics[width=\linewidth]{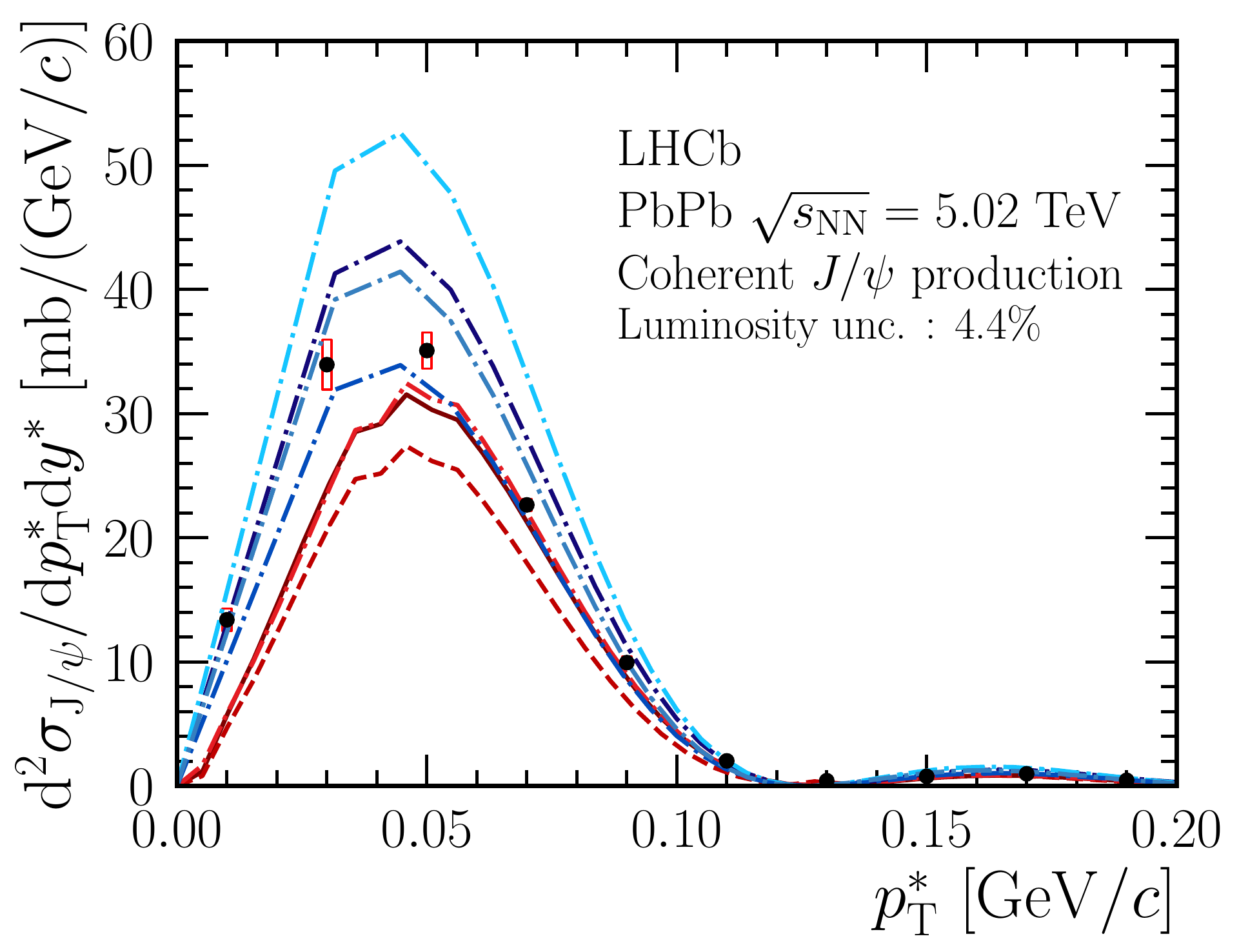}
    \end{minipage}
    \begin{minipage}[t]{0.52\linewidth}
        \centering
        \includegraphics[width=\linewidth]{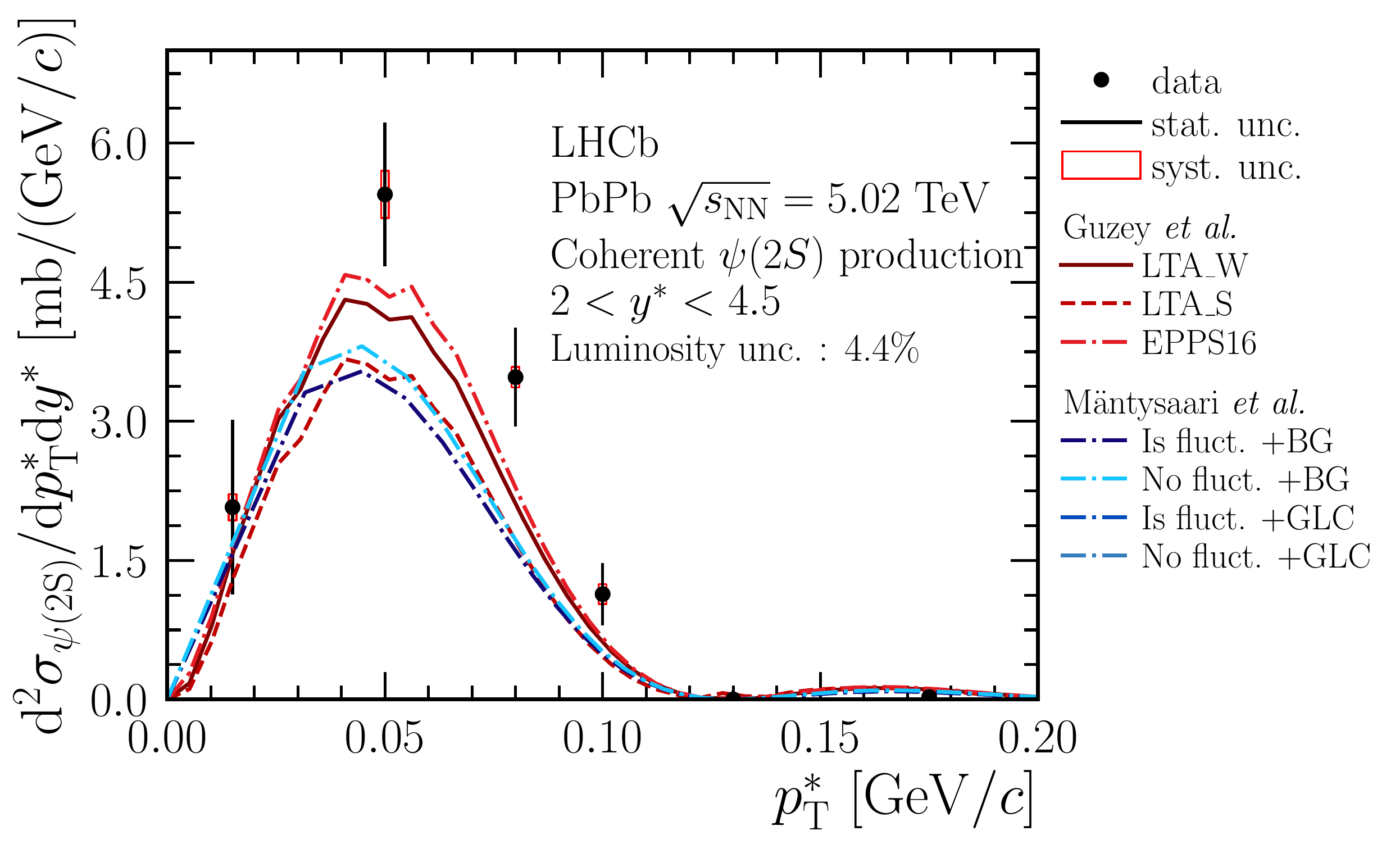}
    \end{minipage}
    \hfil
    \caption{Differential cross-section as a function of $\pt^*$ within the rapidity range $2<y^*<4.5$ for coherent (left) $\jpsi$ and (right) \psitwos production compared to theoretical predictions.}
    \label{fig:theo_pt}
\end{figure}

The results of coherent \jpsi and \psitwos differential cross-sections as a function of rapidity and their ratio are calculated in five rapidity bins and shown in Fig.~\ref{fig:theo_y}.
The differential production cross-sections as a function of transverse momentum are also measured separately and shown in Fig.~\ref{fig:theo_pt}.
We also compared the results between experimental measurements and theoretical predictions~\cite{Guzey_2016,2017access,PhysRevC.84.011902,2018,Kopeliovich:2020has,PhysRevD.96.094027,Gon_alves_2005,20171,Mantysaari:2017dwh,2014arXiv1406.2877L}. 
Also, there are two previous results of coherent \jpsi production in PbPb UPCs by the \lhcb~\cite{Aaij:2775281} and \alice~\cite{2019134926} experiments.
Compared with above measurements, we see this new result is slightly larger than the previous \lhcb result by around 2$\sigma$
and 
more compatible with the \alice measurement. 
\begin{figure}[htbp]
    \centering
    \hfil
    \begin{minipage}[t]{0.49\linewidth}
        \centering
        \includegraphics[width=\linewidth]{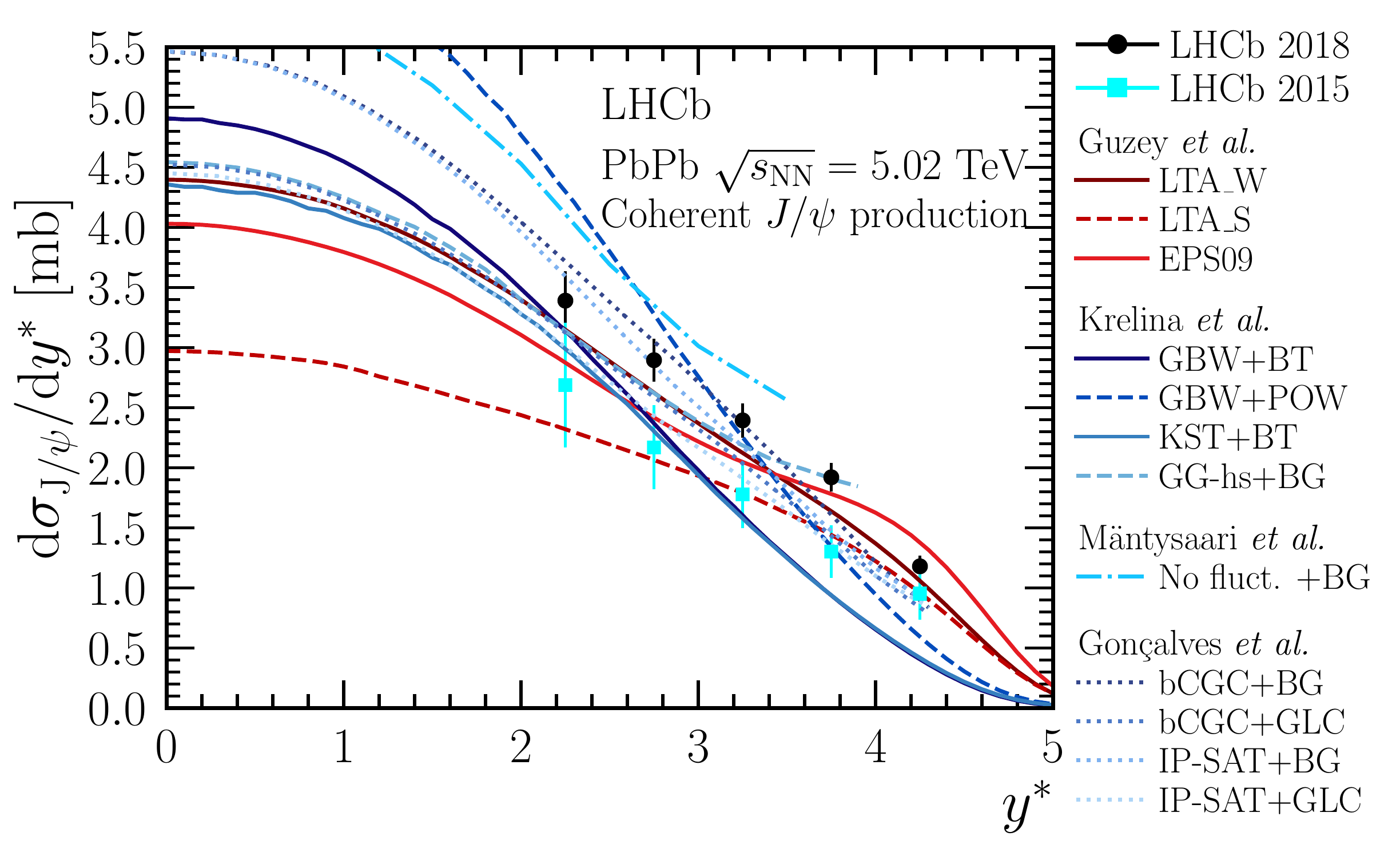}
    \end{minipage}
    \begin{minipage}[t]{0.49\linewidth}
        \centering
        \includegraphics[width=\linewidth]{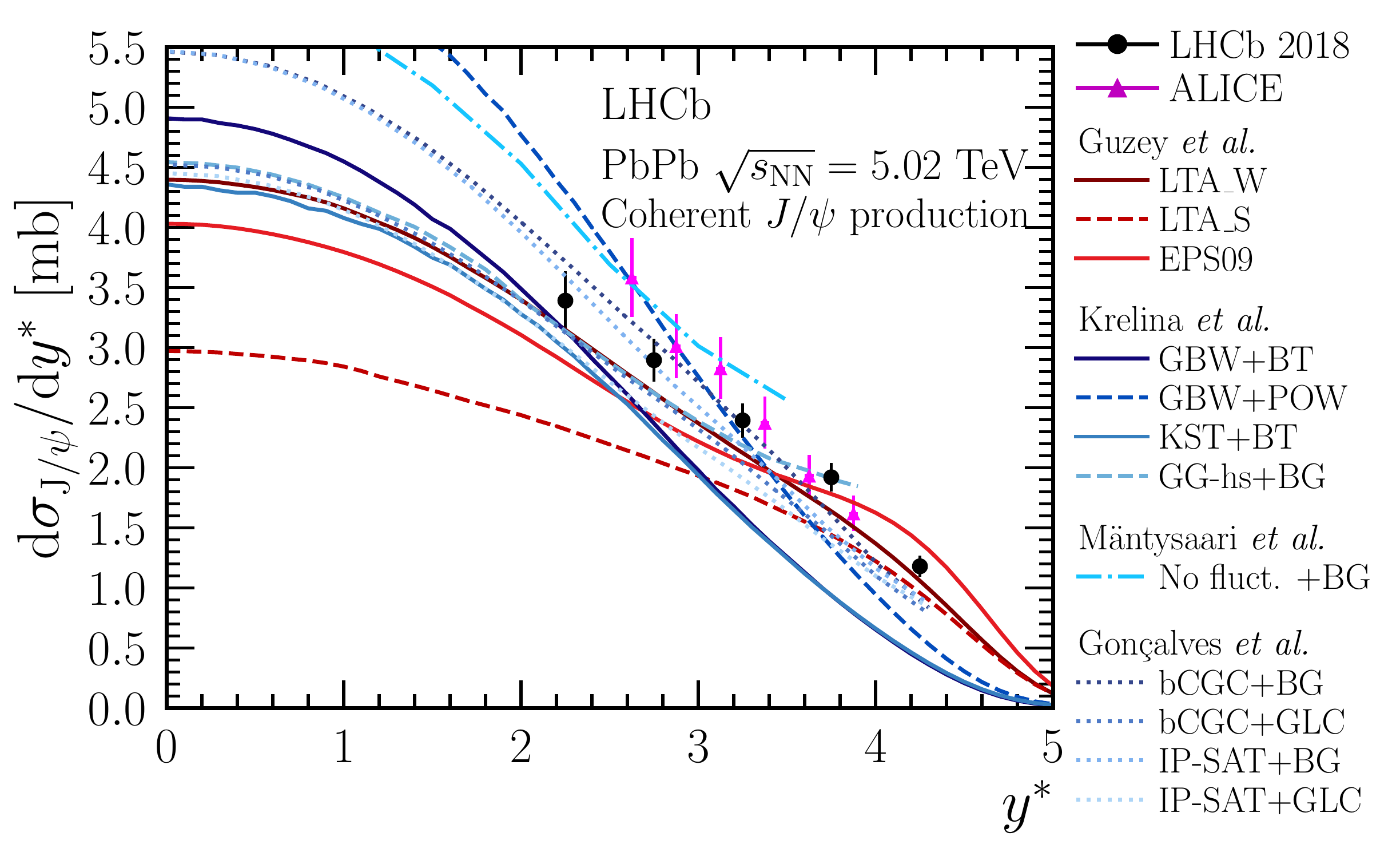}
    \end{minipage}
    \hfil
    \caption{Differential cross-sections as a function of rapidity for coherent \jpsi production compared to the previous measurements by the \lhcb~\cite{Aaij:2775281} and \alice~\cite{2019134926}.
  The measurements are shown as dots and squares, where the uncertainties represent total uncertainties respectively. Then both results are compared to theoretical predictions~\cite{PhysRevC.84.011902,2018,Kopeliovich:2020has,PhysRevD.96.094027,Gon_alves_2005,20171,Guzey_2016,2017access,2014arXiv1406.2877L,Mantysaari:2017dwh}.}
    \label{com}
\end{figure}

\section{Study of J/$\psi$ photo-production in PbPb peripheral collisions at $\sqrt{s_{NN}}= 5\tev$\cite{LHCb:2021hoq}}
\begin{figure}[htbp]
    \centering
    \hfil
    \begin{minipage}[t]{0.49\linewidth}
        \centering
        \includegraphics[width=\linewidth]{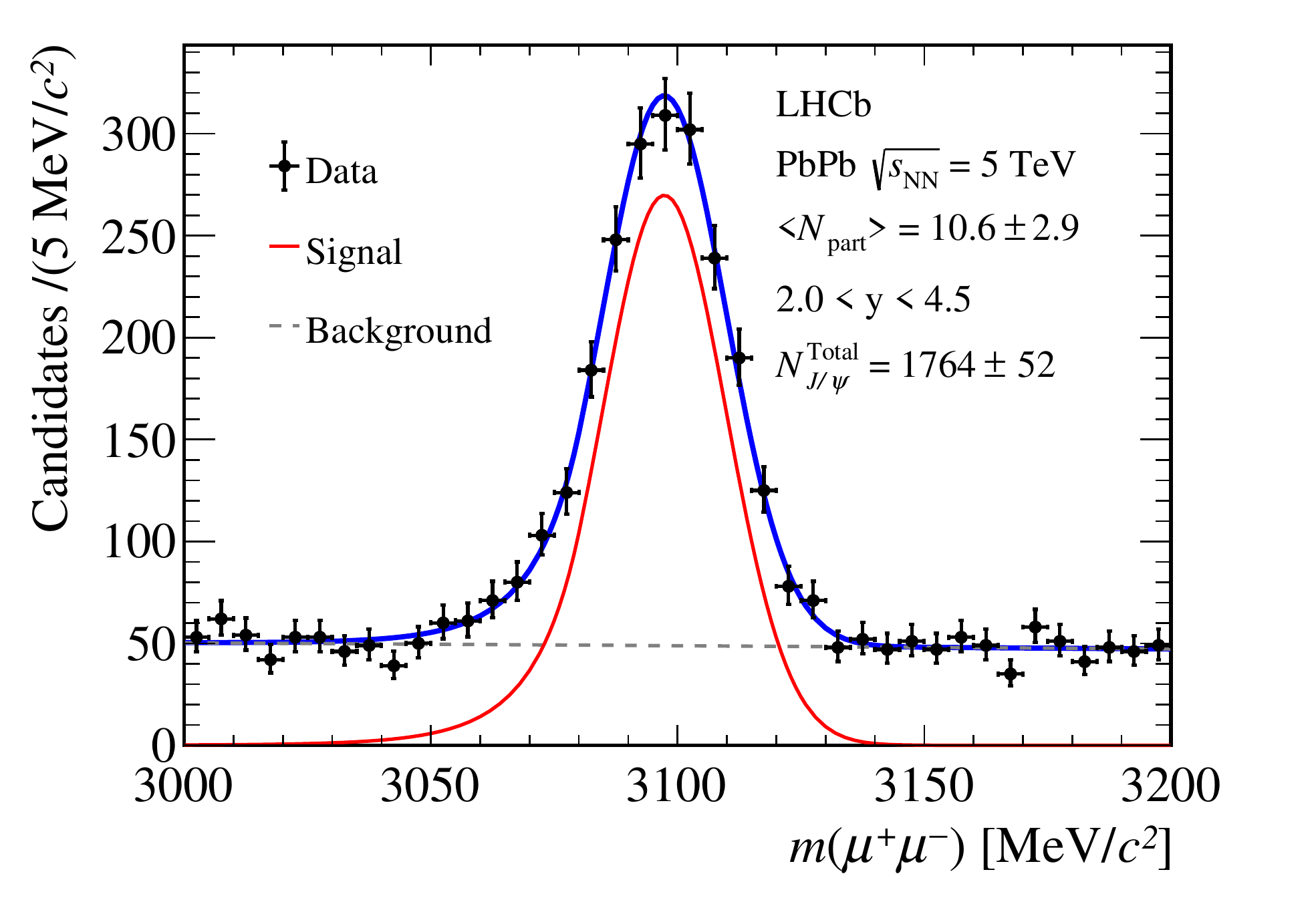}
    \end{minipage}
    \begin{minipage}[t]{0.49\linewidth}
        \centering
        \includegraphics[width=\linewidth]{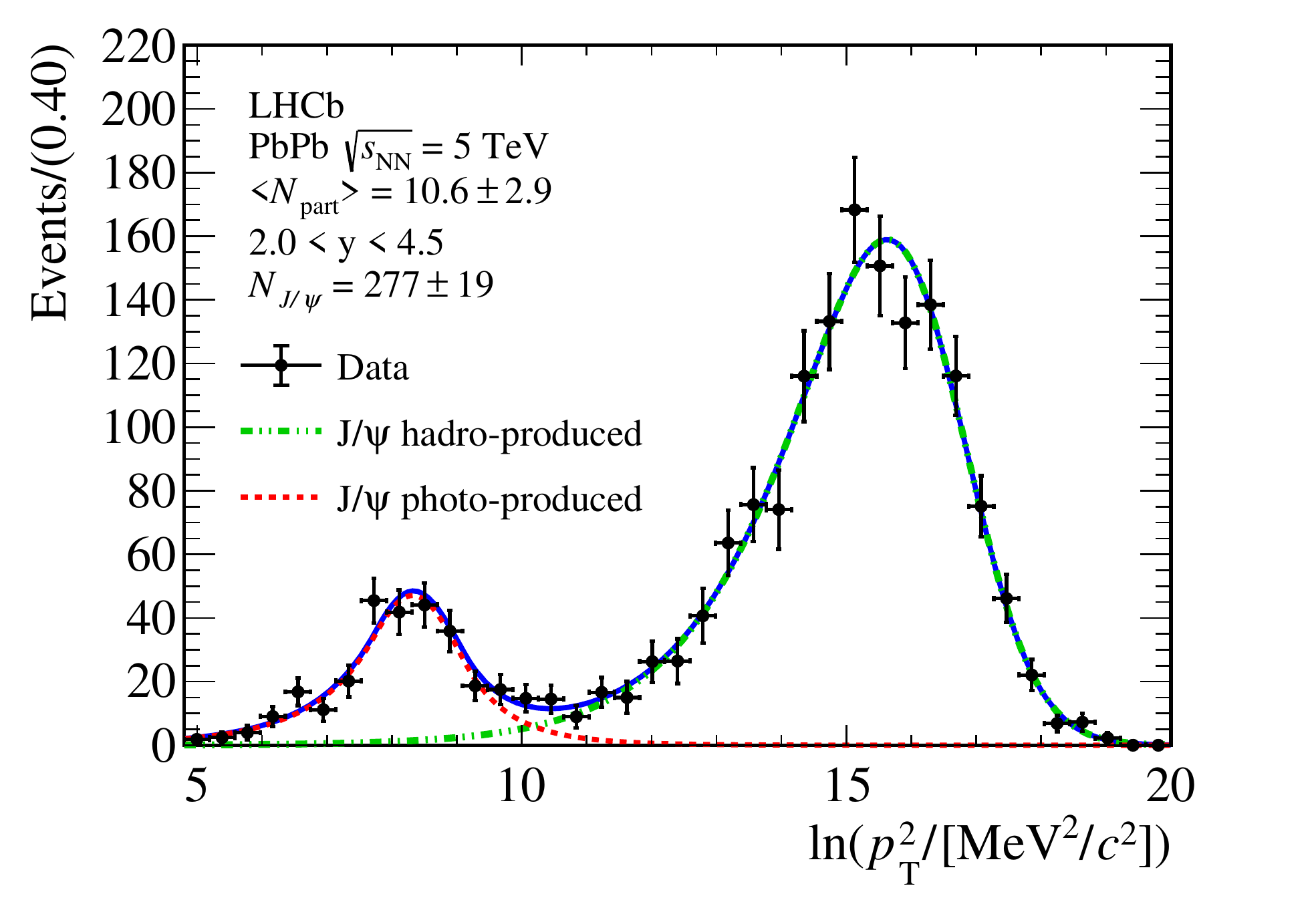}
    \end{minipage}
    \hfil
    \caption{The \logpt2 distribution of dimuon candidates in the $2.0<y^{*}<4.5$ range
    for (left) \jpsi candidates and (right) 
    \psitwos candidates. 
    The data are overlaid with the result of the fit.}
    \label{PC2d}
\end{figure}

In peripheral collisions, the impact parameter b is less than the sum of radii, so there is not only photo-nuclear interaction but also hadronic interaction. 
The photo-produced \jpsi is expected to have a very low transverse momentum of less than 300 \mevc, whereas the hadronic produced \jpsi typically have transverse momentum of about 1-2 \gevc.
Thanks to the high precision of the LHCb experiment, the very low \pt region can be explored and the number of photo-produced \jpsi could be extracted, as shown in Fig.~\ref{PC2d}.

 The differential photo-production yields of \jpsi versus the rapidity, the number of participants in the collision, and the double-differential \jpsi photo-production yields versus transverse momentum as shown in Fig.~\ref{com}, 
 and are compared to theoretical calculations.
One scenario does not consider the destructive effect due to the overlap between the two nuclei, whereas the other takes it into account.
 The two theoretical curves do not show a significant difference because the collisions are peripheral. This nuclear overlapping effect is expected to increase at more central collisions. The trend between \jpsi photo-production measurements and theoretical predictions are consistent in general.

\begin{figure}[ht]
    \centering
    \hfil
    \begin{minipage}[t]{0.32\linewidth}
        \centering
        \includegraphics[width=\linewidth]{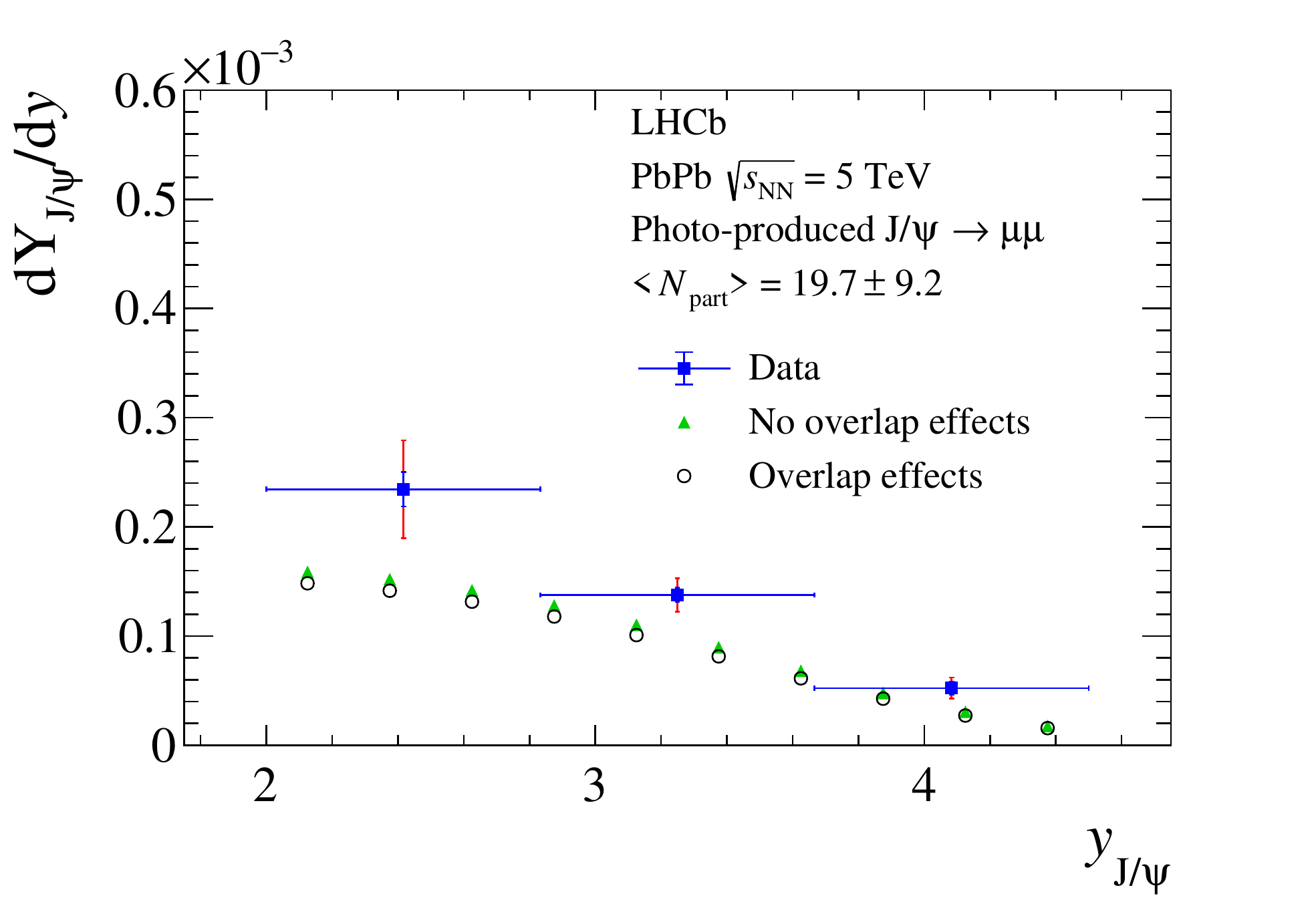}
    \end{minipage}
    \begin{minipage}[t]{0.32\linewidth}
        \centering
        \includegraphics[width=\linewidth]{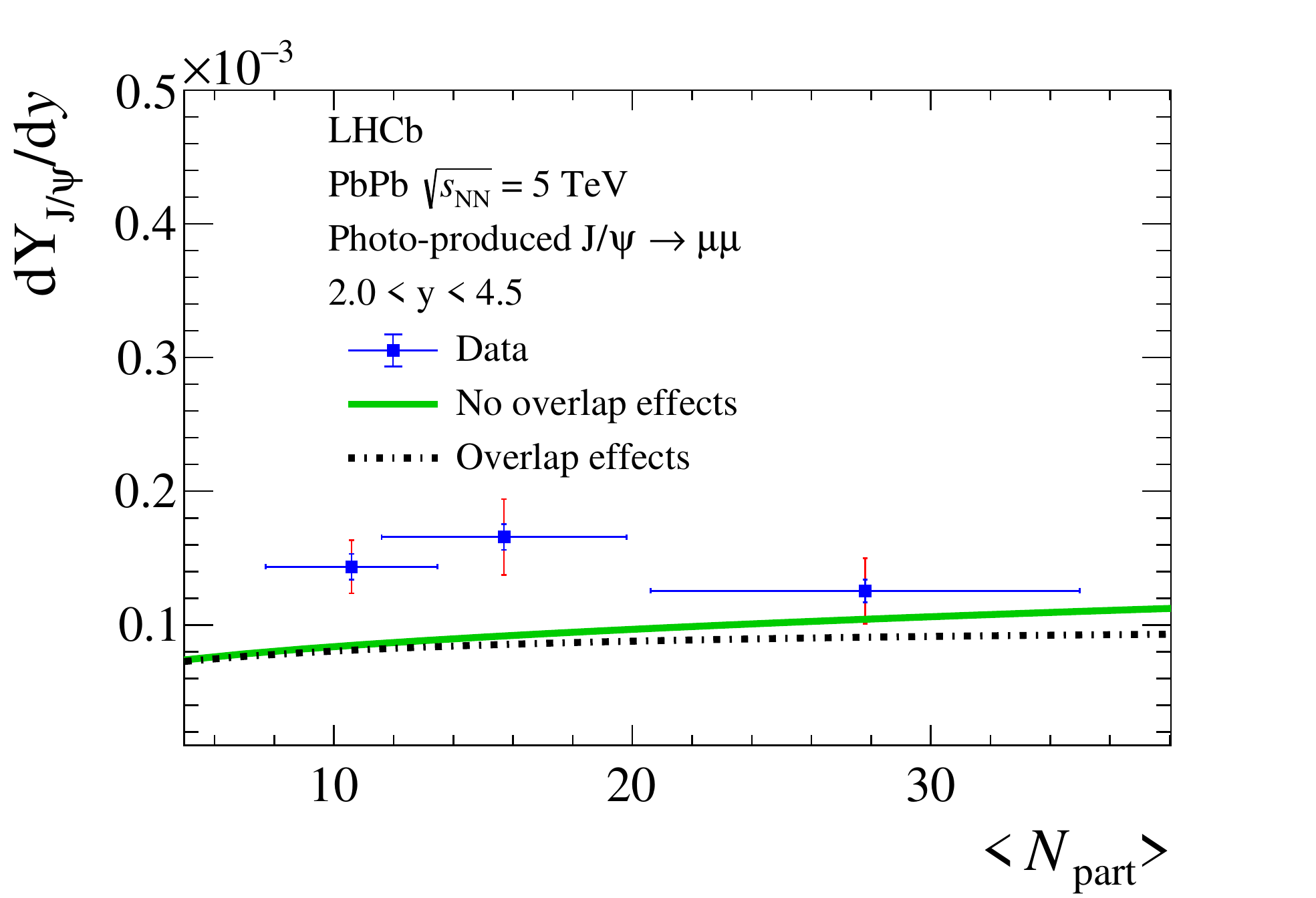}
    \end{minipage}
    \begin{minipage}[t]{0.32\linewidth}
        \centering
        \includegraphics[width=\linewidth]{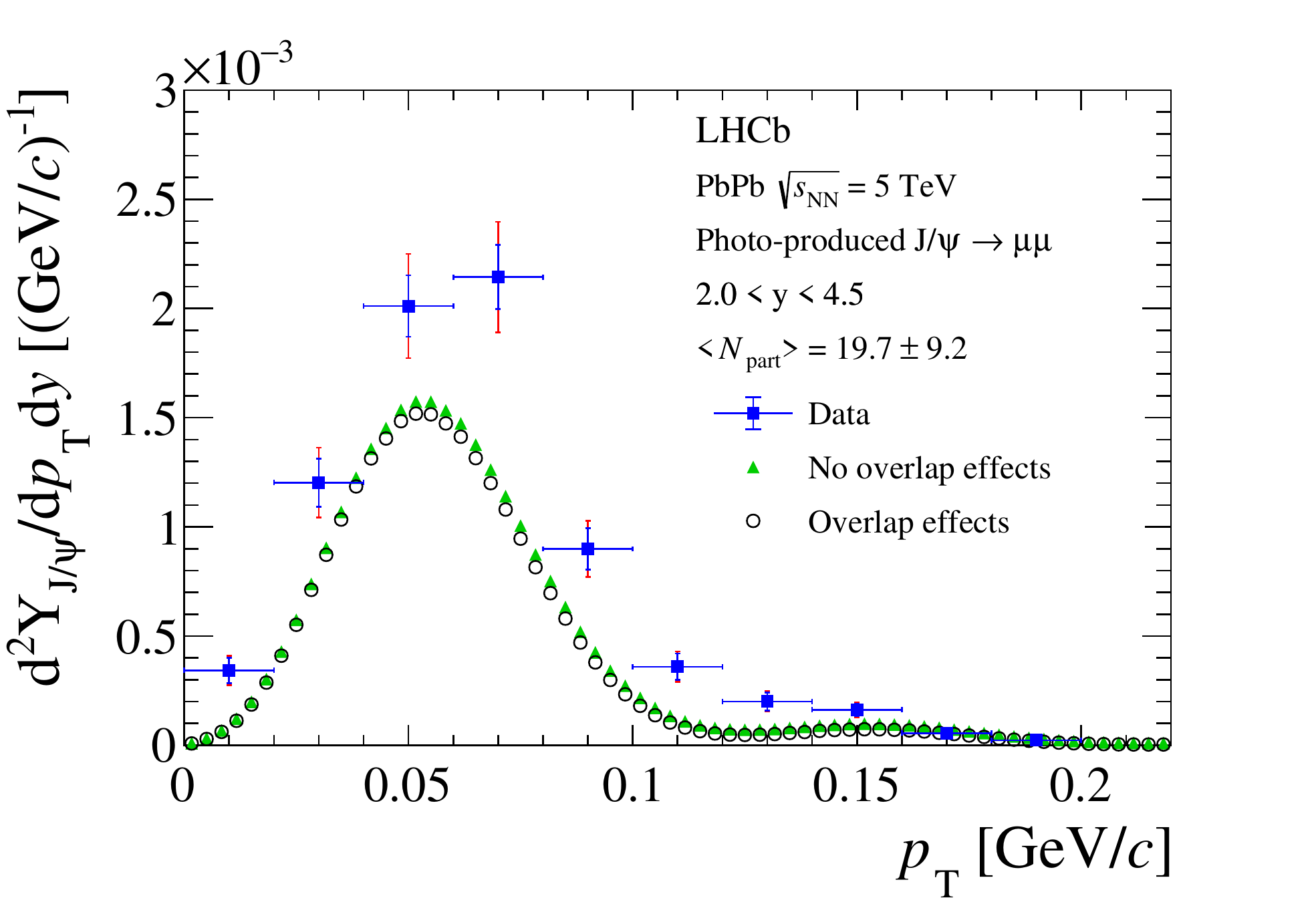}
    \end{minipage}
    \hfil
    \caption{The results of the differential photo-production yields of \jpsi versus the rapidity, the number of participants in the collision and the double-differential \jpsi photo-production yields versus transverse momentum.}
    \label{com}
\end{figure}

\section{Conclusion}
The cross-sections of coherent J/$\psi$, $\psi$(2S) production, and their ratio in UPCs are measured using the 2018 PbPb dataset. It is the first coherent \psitwos measurement in forward rapidity region for UPC at \lhc and the first time to measure the coherent \jpsi and \psitwos production cross-sections versus \pt distribution in PbPb UPCs. The measurement of photo-produced \jpsi mesons in peripheral PbPb collisions using the 2018 dataset is the first result using PbPb hadronic collisions data by the \lhcb and also the most precise coherent photoproduced \jpsi measurement to date. 

\addcontentsline{toc}{section}{References}
\setboolean{inbibliography}{true}
\bibliographystyle{LHCb}
\bibliography{sample,standard,LHCb-PAPER,LHCb-CONF,LHCb-DP,LHCb-TDR}

\ifx\mcitethebibliography\mciteundefinedmacro
\PackageError{LHCb.bst}{mciteplus.sty has not been loaded}
{This bibstyle requires the use of the mciteplus package.}\fi
\providecommand{\href}[2]{#2}
\begin{mcitethebibliography}{10}
\mciteSetBstSublistMode{n}
\mciteSetBstMaxWidthForm{subitem}{\alph{mcitesubitemcount})}
\mciteSetBstSublistLabelBeginEnd{\mcitemaxwidthsubitemform\space}
{\relax}{\relax}

\bibitem{LHCb-DP-2008-001}
LHCb Collaboration, A.~A. Alves~Jr.\ {\em et~al.},
  \ifthenelse{\boolean{articletitles}}{\emph{{The \lhcb detector at the LHC}},
  }{}\href{https://doi.org/10.1088/1748-0221/3/08/S08005}{JINST \textbf{3}
  (2008) S08005}\relax
\mciteBstWouldAddEndPuncttrue
\mciteSetBstMidEndSepPunct{\mcitedefaultmidpunct}
{\mcitedefaultendpunct}{\mcitedefaultseppunct}\relax
\EndOfBibitem
\bibitem{LHCb-DP-2014-002}
LHCb Collaboration, R.~Aaij {\em et~al.},
  \ifthenelse{\boolean{articletitles}}{\emph{{LHCb detector performance}},
  }{}\href{https://doi.org/10.1142/S0217751X15300227}{Int.\ J.\ Mod.\ Phys.\
  \textbf{A30} (2015) 1530022},
  \href{http://arxiv.org/abs/1412.6352}{{\normalfont\ttfamily
  arXiv:1412.6352}}\relax
\mciteBstWouldAddEndPuncttrue
\mciteSetBstMidEndSepPunct{\mcitedefaultmidpunct}
{\mcitedefaultendpunct}{\mcitedefaultseppunct}\relax
\EndOfBibitem
\bibitem{Jones:2015nna}
S.~P. Jones, A.~D. Martin, M.~G. Ryskin, and T.~Teubner,
  \ifthenelse{\boolean{articletitles}}{\emph{{Exclusive $J/\psi$ and $\Upsilon$
  photoproduction and the low $x$ gluon}},
  }{}\href{https://doi.org/10.1088/0954-3899/43/3/035002}{J.\ Phys.\
  \textbf{G43} (2016) 035002},
  \href{http://arxiv.org/abs/1507.06942}{{\normalfont\ttfamily
  arXiv:1507.06942}}\relax
\mciteBstWouldAddEndPuncttrue
\mciteSetBstMidEndSepPunct{\mcitedefaultmidpunct}
{\mcitedefaultendpunct}{\mcitedefaultseppunct}\relax
\EndOfBibitem
\bibitem{LHCb:2022ahs}
LHCb collaboration, \ifthenelse{\boolean{articletitles}}{\emph{{Study of
  coherent charmonium production in ultra-peripheral lead-lead collisions}},
  }{}\href{http://arxiv.org/abs/2206.08221}{{\normalfont\ttfamily
  arXiv:2206.08221}}\relax
\mciteBstWouldAddEndPuncttrue
\mciteSetBstMidEndSepPunct{\mcitedefaultmidpunct}
{\mcitedefaultendpunct}{\mcitedefaultseppunct}\relax
\EndOfBibitem
\bibitem{Bertulani:2005ru}
C.~A. Bertulani, S.~R. Klein, and J.~Nystrand,
  \ifthenelse{\boolean{articletitles}}{\emph{{Physics of ultra-peripheral
  nuclear collisions}},
  }{}\href{https://doi.org/10.1146/annurev.nucl.55.090704.151526}{Ann.\ Rev.\
  Nucl.\ Part.\ Sci.\  \textbf{55} (2005) 271},
  \href{http://arxiv.org/abs/nucl-ex/0502005}{{\normalfont\ttfamily
  arXiv:nucl-ex/0502005}}\relax
\mciteBstWouldAddEndPuncttrue
\mciteSetBstMidEndSepPunct{\mcitedefaultmidpunct}
{\mcitedefaultendpunct}{\mcitedefaultseppunct}\relax
\EndOfBibitem
\bibitem{Guzey_2016}
V.~Guzey, E.~Kryshen, and M.~Zhalov,
  \ifthenelse{\boolean{articletitles}}{\emph{{Coherent photoproduction of
  vector mesons in ultraperipheral heavy ion collisions: Update for run2 at the
  CERN Large Hadron Collider}},
  }{}\href{https://doi.org/10.1103/PhysRevC.93.055206}{Phys.\ Rev.\
  \textbf{C93} (2016) 055206},
  \href{http://arxiv.org/abs/1602.01456}{{\normalfont\ttfamily
  arXiv:1602.01456}}\relax
\mciteBstWouldAddEndPuncttrue
\mciteSetBstMidEndSepPunct{\mcitedefaultmidpunct}
{\mcitedefaultendpunct}{\mcitedefaultseppunct}\relax
\EndOfBibitem
\bibitem{2017access}
V.~Guzey, M.~Strikman, and M.~Zhalov,
  \ifthenelse{\boolean{articletitles}}{\emph{{Accessing transverse nucleon and
  gluon distributions in heavy nuclei using coherent vector meson
  photoproduction at high energies in ion ultraperipheral collisions}},
  }{}\href{https://doi.org/10.1103/PhysRevC.95.025204}{Phys.\ Rev.\
  \textbf{C95} (2017) 025204},
  \href{http://arxiv.org/abs/1611.05471}{{\normalfont\ttfamily
  arXiv:1611.05471}}\relax
\mciteBstWouldAddEndPuncttrue
\mciteSetBstMidEndSepPunct{\mcitedefaultmidpunct}
{\mcitedefaultendpunct}{\mcitedefaultseppunct}\relax
\EndOfBibitem
\bibitem{PhysRevC.84.011902}
V.~P. Gon\ifmmode~\mbox{\c{c}}\else \c{c}\fi{}alves and M.~V.~T. Machado,
  \ifthenelse{\boolean{articletitles}}{\emph{{Vector meson production in
  coherent hadronic interactions: an update on predictions for RHIC and LHC}},
  }{}\href{https://doi.org/10.1103/PhysRevC.84.011902}{Phys.\ Rev.\
  \textbf{C84} (2011) 011902},
  \href{http://arxiv.org/abs/1106.3036}{{\normalfont\ttfamily
  arXiv:1106.3036}}\relax
\mciteBstWouldAddEndPuncttrue
\mciteSetBstMidEndSepPunct{\mcitedefaultmidpunct}
{\mcitedefaultendpunct}{\mcitedefaultseppunct}\relax
\EndOfBibitem
\bibitem{2018}
J.~Cepila, J.~G. Contreras, and M.~Krelina,
  \ifthenelse{\boolean{articletitles}}{\emph{{Coherent and incoherent
  $\mathrm{J/}\psi$ photonuclear production in an energy-dependent hot-spot
  model}}, }{}\href{https://doi.org/10.1103/PhysRevC.97.024901}{Phys.\ Rev.\
  \textbf{C97} (2018) 024901},
  \href{http://arxiv.org/abs/1711.01855}{{\normalfont\ttfamily
  arXiv:1711.01855}}\relax
\mciteBstWouldAddEndPuncttrue
\mciteSetBstMidEndSepPunct{\mcitedefaultmidpunct}
{\mcitedefaultendpunct}{\mcitedefaultseppunct}\relax
\EndOfBibitem
\bibitem{Kopeliovich:2020has}
B.~Z. Kopeliovich, M.~Krelina, J.~Nemchik, and I.~K. Potashnikova,
  \ifthenelse{\boolean{articletitles}}{\emph{{Heavy quarkonium production in
  ultraperipheral nuclear collisions}},
  }{}\href{http://arxiv.org/abs/2008.05116}{{\normalfont\ttfamily
  arXiv:2008.05116}}\relax
\mciteBstWouldAddEndPuncttrue
\mciteSetBstMidEndSepPunct{\mcitedefaultmidpunct}
{\mcitedefaultendpunct}{\mcitedefaultseppunct}\relax
\EndOfBibitem
\bibitem{PhysRevD.96.094027}
V.~P. Gon\ifmmode~\mbox{\c{c}}\else \c{c}\fi{}alves {\em et~al.},
  \ifthenelse{\boolean{articletitles}}{\emph{Color dipole predictions for the
  exclusive vector meson photoproduction in $pp$, $p\mathrm{Pb}$, and
  $\mathrm{Pb}\mathrm{Pb}$ collisions at run2 \lhc energies},
  }{}\href{https://doi.org/10.1103/PhysRevD.96.094027}{Phys.\ Rev.\
  \textbf{D96} (2017) 094027},
  \href{http://arxiv.org/abs/1710.10070}{{\normalfont\ttfamily
  arXiv:1710.10070}}\relax
\mciteBstWouldAddEndPuncttrue
\mciteSetBstMidEndSepPunct{\mcitedefaultmidpunct}
{\mcitedefaultendpunct}{\mcitedefaultseppunct}\relax
\EndOfBibitem
\bibitem{Gon_alves_2005}
V.~P. Gon\ifmmode~\mbox{\c{c}}\else \c{c}\fi{}alves and M.~V.~T. Machado,
  \ifthenelse{\boolean{articletitles}}{\emph{{The QCD pomeron in
  ultraperipheral heavy ion collisions: IV. Photonuclear production of vector
  mesons}}, }{}\href{https://doi.org/10.1140/epjc/s2005-02175-3}{Eur.\ Phys.\
  J.\  \textbf{C40} (2005) 519},
  \href{http://arxiv.org/abs/hep-ph/0501099}{{\normalfont\ttfamily
  arXiv:hep-ph/0501099}}\relax
\mciteBstWouldAddEndPuncttrue
\mciteSetBstMidEndSepPunct{\mcitedefaultmidpunct}
{\mcitedefaultendpunct}{\mcitedefaultseppunct}\relax
\EndOfBibitem
\bibitem{20171}
H.~Kowalski, L.~Motyka, and G.~Watt,
  \ifthenelse{\boolean{articletitles}}{\emph{{Exclusive diffractive processes
  at HERA within the dipole picture}},
  }{}\href{https://doi.org/10.1103/PhysRevD.74.074016}{Phys.\ Rev.\
  \textbf{D74} (2006) 074016},
  \href{http://arxiv.org/abs/hep-ph/0606272}{{\normalfont\ttfamily
  arXiv:hep-ph/0606272}}\relax
\mciteBstWouldAddEndPuncttrue
\mciteSetBstMidEndSepPunct{\mcitedefaultmidpunct}
{\mcitedefaultendpunct}{\mcitedefaultseppunct}\relax
\EndOfBibitem
\bibitem{Mantysaari:2017dwh}
H.~M\"antysaari and B.~Schenke,
  \ifthenelse{\boolean{articletitles}}{\emph{{Probing subnucleon scale
  fluctuations in ultraperipheral heavy ion collisions}},
  }{}\href{https://doi.org/10.1016/j.physletb.2017.07.063}{Phys.\ Lett.\
  \textbf{B772} (2017) 832},
  \href{http://arxiv.org/abs/1703.09256}{{\normalfont\ttfamily
  arXiv:1703.09256}}\relax
\mciteBstWouldAddEndPuncttrue
\mciteSetBstMidEndSepPunct{\mcitedefaultmidpunct}
{\mcitedefaultendpunct}{\mcitedefaultseppunct}\relax
\EndOfBibitem
\bibitem{2014arXiv1406.2877L}
T.~Lappi and H.~M\"antysaari,
  \ifthenelse{\boolean{articletitles}}{\emph{{Diffractive vector meson
  production in ultraperipheral heavy ion collisions from the color glass
  condensate}}, }{}\href{https://doi.org/10.22323/1.203.0069}{PoS
  \textbf{DIS2014} (2014) 069},
  \href{http://arxiv.org/abs/1406.2877}{{\normalfont\ttfamily
  arXiv:1406.2877}}\relax
\mciteBstWouldAddEndPuncttrue
\mciteSetBstMidEndSepPunct{\mcitedefaultmidpunct}
{\mcitedefaultendpunct}{\mcitedefaultseppunct}\relax
\EndOfBibitem
\bibitem{Aaij:2775281}
LHCb collaboration, R.~Aaij {\em et~al.},
  \ifthenelse{\boolean{articletitles}}{\emph{{Study of coherent $J/\psi$
  production in lead-lead collisions at $\sqrt{s_{NN}} = 5 TeV$}},
  }{}\href{http://arxiv.org/abs/2107.03223}{{\normalfont\ttfamily
  arXiv:2107.03223}}\relax
\mciteBstWouldAddEndPuncttrue
\mciteSetBstMidEndSepPunct{\mcitedefaultmidpunct}
{\mcitedefaultendpunct}{\mcitedefaultseppunct}\relax
\EndOfBibitem
\bibitem{2019134926}
ALICE collaboration, S.~Acharya {\em et~al.},
  \ifthenelse{\boolean{articletitles}}{\emph{{Coherent J/$\psi$ photoproduction
  at forward rapidity in ultra-peripheral Pb-Pb collisions at
  $\sqrt{s_{\rm{NN}}}=5.02$ TeV}},
  }{}\href{https://doi.org/10.1016/j.physletb.2019.134926}{Phys.\ Lett.\
  \textbf{B798} (2019) 134926},
  \href{http://arxiv.org/abs/1904.06272}{{\normalfont\ttfamily
  arXiv:1904.06272}}\relax
\mciteBstWouldAddEndPuncttrue
\mciteSetBstMidEndSepPunct{\mcitedefaultmidpunct}
{\mcitedefaultendpunct}{\mcitedefaultseppunct}\relax
\EndOfBibitem
\bibitem{LHCb:2021hoq}
LHCb collaboration, R.~Aaij {\em et~al.},
  \ifthenelse{\boolean{articletitles}}{\emph{{$J/\psi$ photoproduction in Pb-Pb
  peripheral collisions at $\sqrt {s_{NN}}$= 5 TeV}},
  }{}\href{https://doi.org/10.1103/PhysRevC.105.L032201}{Phys.\ Rev.\ C
  \textbf{105} (2022) L032201},
  \href{http://arxiv.org/abs/2108.02681}{{\normalfont\ttfamily
  arXiv:2108.02681}}\relax
\mciteBstWouldAddEndPuncttrue
\mciteSetBstMidEndSepPunct{\mcitedefaultmidpunct}
{\mcitedefaultendpunct}{\mcitedefaultseppunct}\relax
\EndOfBibitem
\end{mcitethebibliography}

\end{document}